\documentclass[prl,aps,preprint,showpacs,superscriptaddress,floatfix]{revtex4-1}
\usepackage{amssymb}
\usepackage{graphicx}
\usepackage{float}
\usepackage{amsmath}

\begin{document}

\title[High-field Fermi surface studies of an electron-doped cuprate
superconductor]{Fermi Surface of the Electron-doped Cuprate Superconductor
$\mathbf{Nd_{2-x}Ce_xCuO_{4}}$ Probed by High-Field Magnetotransport}

\author{M. V. Kartsovnik$^1$, T. Helm$^1$, C. Putzke$^2$, F. Wolff-Fabris$^2$,
I. Sheikin$^3$, S. Lepault$^4$, C. Proust$^4$, D. Vignolles$^4$, N.
Bittner$^1$, W. Biberacher$^1$, A. Erb$^1$, J. Wosnitza$^2$, and R.
Gross$^{1,5}$}

\affiliation{$^1$ Walther-Mei{\ss}ner-Institut, Bayerische Akademie der
Wissenschaften, Walther-Mei{\ss}ner-Str. 8, D-85748 Garching, Germany}
\affiliation{$^2$ Hochfeld-Magnetlabor Dresden, Forschungszentrum
Dresden-Rossendorf, Bautzner Landstr. 400, D-01328 Dresden, Germany}
\affiliation{$^3$ Laboratoire National des Champs Magn\'{e}tiques Intenses,
CNRS, 25 rue des Martyrs, B.P. 166, F-38042 Grenoble, France}
\affiliation{$^4$ Laboratoire National des Champs Magn\'{e}tiques Intenses,
(CNRS, INSA, UJF, UPS), F-31400 Toulouse, France}
\affiliation{$^5$ Physik-Department, Technische Universit\"{a}t M\"{u}nchen,
James-Franck-Str., D-85748 Garching, Germany}
\email{Mark.Kartsovnik@wmi.badw.de}

\begin{abstract}
We report on the study of the Fermi surface of the electron-doped cuprate
superconductor Nd$_{2-x}$Ce$_x$CuO$_{4}$ by measuring the interlayer
magnetoresistance as a function of the strength and orientation of the applied
magnetic field. We performed experiments in both steady and pulsed magnetic
fields on high-quality single crystals with Ce concentrations of $x=0.13$ to $0.17$.
In the overdoped regime of $x > 0.15$ we found both semiclassical
angle-dependent magnetoresistance oscillations (AMRO) and Shubnikov-de Haas
(SdH) oscillations. The combined AMRO and SdH data clearly show that the
appearance of fast SdH oscillations in strongly overdoped samples is caused by
magnetic breakdown. This observation provides clear evidence for a
reconstructed multiply-connected Fermi surface up to the very end of the overdoped
regime at $x\simeq 0.17$. The strength of the superlattice potential
responsible for the reconstructed Fermi surface is found to decrease with
increasing doping level and likely vanishes at the same carrier concentration
as superconductivity, suggesting a close relation between translational
symmetry breaking and superconducting pairing. A detailed analysis of the
high-resolution SdH data allowed us to determine the effective cyclotron mass
and Dingle temperature, as well as to estimate the magnetic breakdown field
in the overdoped regime.
\end{abstract}

\pacs{74.72.Ek, %Electron-doped
      71.18.+y, %Fermi surface: calculations and measurements; effective mass, g factor
      72.15.Gd, %Galvanomagnetic and other magnetotransport effects
      74.25.Jb} %Electronic structure (photoemission, etc.)

\maketitle

\section{Introduction}
\label{sec:intro}

Electronic correlations and the resulting ordering instabilities are central
issues in the long-standing problem of high-temperature superconductivity in
copper oxides. To elucidate them, the exact knowledge of the Fermi surface and
its evolution with doping is of crucial importance. High-field magnetotransport
is known as one of the most powerful tools for studying Fermi surfaces of
conventional metals~\cite{crac73,ashc76}. It has recently proved
very efficient also in the case of cuprate superconductors. A
breakthrough in the Fermiology of hole-doped cuprates was made by the
observation of semiclassical angle-dependent magnetoresistance oscillations
(AMRO)~\cite{huss03,huss06} and quantum oscillations of the resistance, the
Shubnikov-de Haas (SdH) effect~\cite{doir07,yell08,bang08,vign08}.

\begin{figure}[tb]
\centering
\includegraphics[width=0.9\columnwidth]{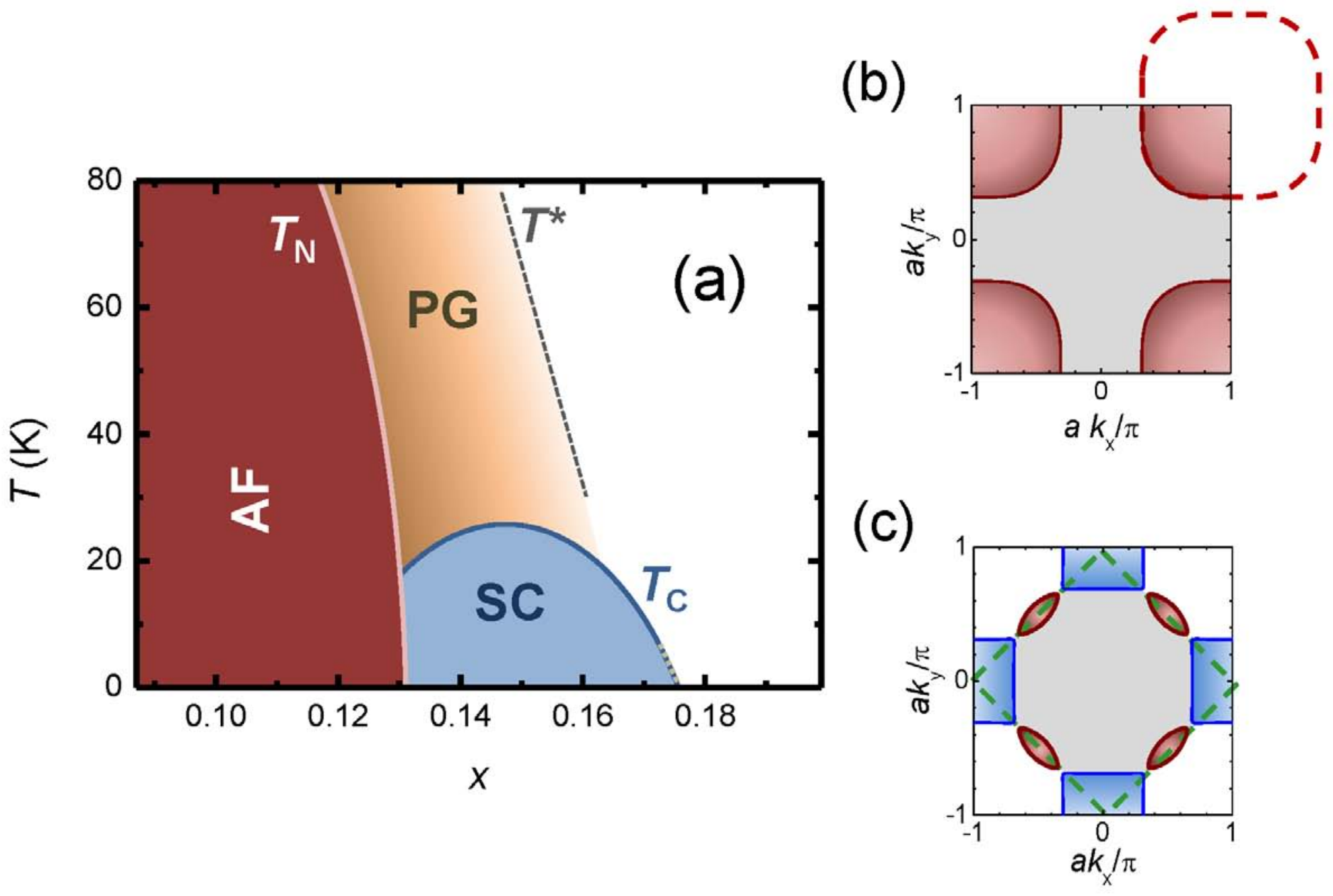}
\caption{
(a) Schematic phase diagram of Nd$_{2-x}$Ce$_x$CuO$_{4}$ including the
normal conducting, the pseudogap (PG), the antiferromagnetic insulating (AF),
and the superconducting (SC) state together with the corresponding transition
temperatures $T^{\ast}$, $T_{\rm N}$, and $T_{\rm c}$ depending on the Ce
concentration $x$. (b) The Fermi surface of NCCO in the absence of a
superlattice potential is represented by a single hole-like cylinder centered
at the corner of the Brillouin zone and slightly warped in the direction
perpendicular to CuO$_2$ layers. In a strong magnetic field charge carriers
move on large closed orbits (red dashed line) encircling the area corresponding
to the fast SdH oscillations. (c) The Fermi surface, reconstructed due to a
$(\pi/a,\pi/a)$ superlattice potential, consists of electron (blue) and
hole (dark red) pockets. The size of the small hole pockets is consistent with
the frequency of slow oscillations observed for the optimal, $x=0.15$, and
slightly overdoped, $x=0.16$, compositions~\cite{helm09}.}
 \label{fig1}
\end{figure}

On the strongly overdoped side of the phase diagram of hole-doped cuprate
superconductors, both the semiclassical AMRO~\cite{huss03} and quantum
oscillations of the resistance~\cite{vign08} have been found for the compound
Tl$_2$Ba$_2$CuO$_{6+\delta}$. These experiments provided evidence for a large
cylindrical Fermi surface, as expected from band-structure
calculations~\cite{ande95,sing92} and angle-resolved photoemission spectroscopy
(ARPES)~\cite{plat05}. In contrast, for underdoped
YBa$_2$Cu$_3$O$_{6.5}$~\cite{doir07,seba08,jaud08} and
YBa$_2$Cu$_4$O$_8$~\cite{yell08,bang08} slow SdH and de Haas-van Alphen (dHvA)
oscillations were found, indicating a reconstruction of the Fermi
surface. These observations reveal substantial disagreements with ARPES
results~\cite{dama03} and are controversially interpreted at present, see e.g.
Refs.~\cite{seba10,rams10,jia09,seba08,lee08,mill07,lebo07,kaul08,dimo08,alex08,varm09,pere10}.
The electron-doped cuprates $Ln_{2-x}$Ce$_x$CuO$_{4}$
($Ln=$ Nd, Pr, Sm) have a number of advantages for high-field Fermi surface
studies, as compared to hole-doped cuprates. Due to their lower critical
fields, superconductivity can easily be suppressed and the normal state is
accessed for any doping level even at the lowest temperatures by applying
a magnetic field $B \gtrsim 10$\,T (perpendicular to CuO$_2$ layers).
Moreover, the Fermi surface is expected to be simple: there are
neither CuO chains nor any bilayer potential and the magnetic superstructure,
established in underdoped compounds, is commensurate. Another important
advantage is that, in contrast to most of the hole-doped cuprates, the entire
doping range, from the undoped insulating up to the strongly overdoped metallic
(superconducting) phase, can be covered using one and the same compound with
just slightly different Ce concentrations [see Fig.~\ref{fig1}(a)].

We have recently reported on SdH oscillations in
Nd$_{2-x}$Ce$_x$CuO$_{4}$ (NCCO) single crystals with Ce concentrations
corresponding to nearly optimal doping ($x=0.15$) and to overdoped compositions
($x=0.16$ and 0.17)~\cite{helm09}. For the highest doping level, $x = 0.17$,
corresponding to the very end of the superconducting region on the overdoped
side, fast SdH oscillations were found. In complete analogy with the
hole-overdoped Tl$_2$Ba$_2$CuO$_{6+\delta}$ (Tl2201) compound, the
oscillation frequency, $F_{\mathrm{fast}} \approx 11$~kT, reveals a large
cyclotron orbit on the cylindrical Fermi surface centered at the corner of the
Brillouin zone, as shown in Fig.~\ref{fig1}(b), and occupying $\approx 41\%$ of
the first Brillouin zone. This result is in full agreement with
band-structure calculations~\cite{mass89} and ARPES
experiments~\cite{armi02,mats07}. However, on crystals with the Ce
concentration reduced by just 1\% no fast oscillations have been found.
Instead, oscillations with a much lower frequency have been observed for
$x=0.16$ and 0.15. These slow oscillations indicate a reconstruction of the Fermi
surface caused by a translational symmetry breaking in the electronic system. Most
remarkably, the transformation to a reconstructed Fermi surface occurs deep
in the overdoped regime. While this result is in good agreement with the
Hall effect measurements on NCCO crystals \cite{lamb08} and
Pr$_{2-x}$Ce$_x$CuO$_{4}$
thin films~\cite{daga04}, it is obviously inconsistent with the conclusions of
ARPES~\cite{armi02,mats07}  and neutron-scattering studies~\cite{mats07}.

An important issue to be clarified is the exact origin of the Fermi surface
reconstruction which was revealed by the SdH oscillations in our overdoped NCCO
samples. It was proposed~\cite{helm09,chak10} to be caused by a $(\pi/a,\pi/a)$
superlattice potential ($a$ is the lattice constant within the CuO$_2$ plane)
which is known to exist in undoped and underdoped NCCO~\cite{armi02,kusk02}.
Indeed, the frequency of the slow SdH oscillations, $F_{\mathrm{slow}} \approx
300$\,T, is consistent with the size of small hole pockets, which should be
formed around $(\pm\pi/2a,\pm\pi/2a)$ due to such ordering [see Fig.~\ref{fig1}(c)].
On the other hand, no indication of electron pockets, required by the proposed reconstructed
Fermi surface topology, were found in the SdH spectra. Obviously, additional
work is needed for further verifying the reconstruction scenario. For example,
it would be desirable to determine not only the size but also the shape of the
Fermi pockets. Towards this end, the study of semiclassical AMRO is a very
efficient tool. AMRO have been widely used for mapping in-plane Fermi surfaces
of organic conductors~\cite{kart04} and other layered systems such as
Sr$_2$RuO$_4$~\cite{berg03} and intercalated graphite~\cite{baxe98}. This effect
has a geometrical origin and is directly related to the shape of a weakly warped
cylindrical Fermi surface~\cite{huss03,yama89,kart92,grig10}. As mentioned above,
AMRO already have been observed in hole-overdoped Tl$_2$Ba$_2$CuO$_{6+\delta}$ (Tl2201)
samples~\cite{huss03,huss06} and successfully used for extracting the shape of
the three-dimensional Fermi surface as well as for evaluating the scattering
anisotropy.

Very recently, we have also found features characteristic of AMRO in the
angle-dependent interlayer magnetoresistance of overdoped NCCO~\cite{helm10-}.
Although the magnitude of these features observed in applied magnetic fields up
to 28\,T was too low for a reliable quantitative analysis, they provided an
important argument for the existence of magnetic-breakdown orbits on the Fermi
surface. This magnetic breakdown scenario was further supported by the
observation of two frequencies in the SdH spectrum obtained for strongly
overdoped ($x=0.17$) NCCO crystals~\cite{helm10-}.

Here, we present new data on the interlayer magnetoresistance of NCCO single
crystals studied as a function of the orientation and strength of the applied
magnetic field at Ce concentrations between $x=0.16$ and 0.17, corresponding to
the overdoped regime of the phase diagram [see Fig.~\ref{fig1}(a)]. Our studies
confirm the existence of AMRO and reveal magnetic-breakdown quantum oscillations
at compositions down to $x=0.16$.

\section{Experimental Techniques}
\label{sec:exp}

Single crystals of NCCO were grown by the traveling solvent floating zone
(TSFZ) method and thermally treated in pure argon to remove interstitial oxygen and
release internal strain. The crystals were thoroughly tested by x-ray
diffraction, magnetic and resistive measurements. The best samples, with the
lowest doping inhomogeneity (typically within $0.25\%$) and the largest
(high-to-low temperature) resistance ratios, see Fig.~2(a), were selected for the
high-field experiments. Details of crystal preparation and characterization are
presented elsewhere~\cite{lamb10}.

\begin{figure}[tb]
\centering
\includegraphics[width=0.9\columnwidth]{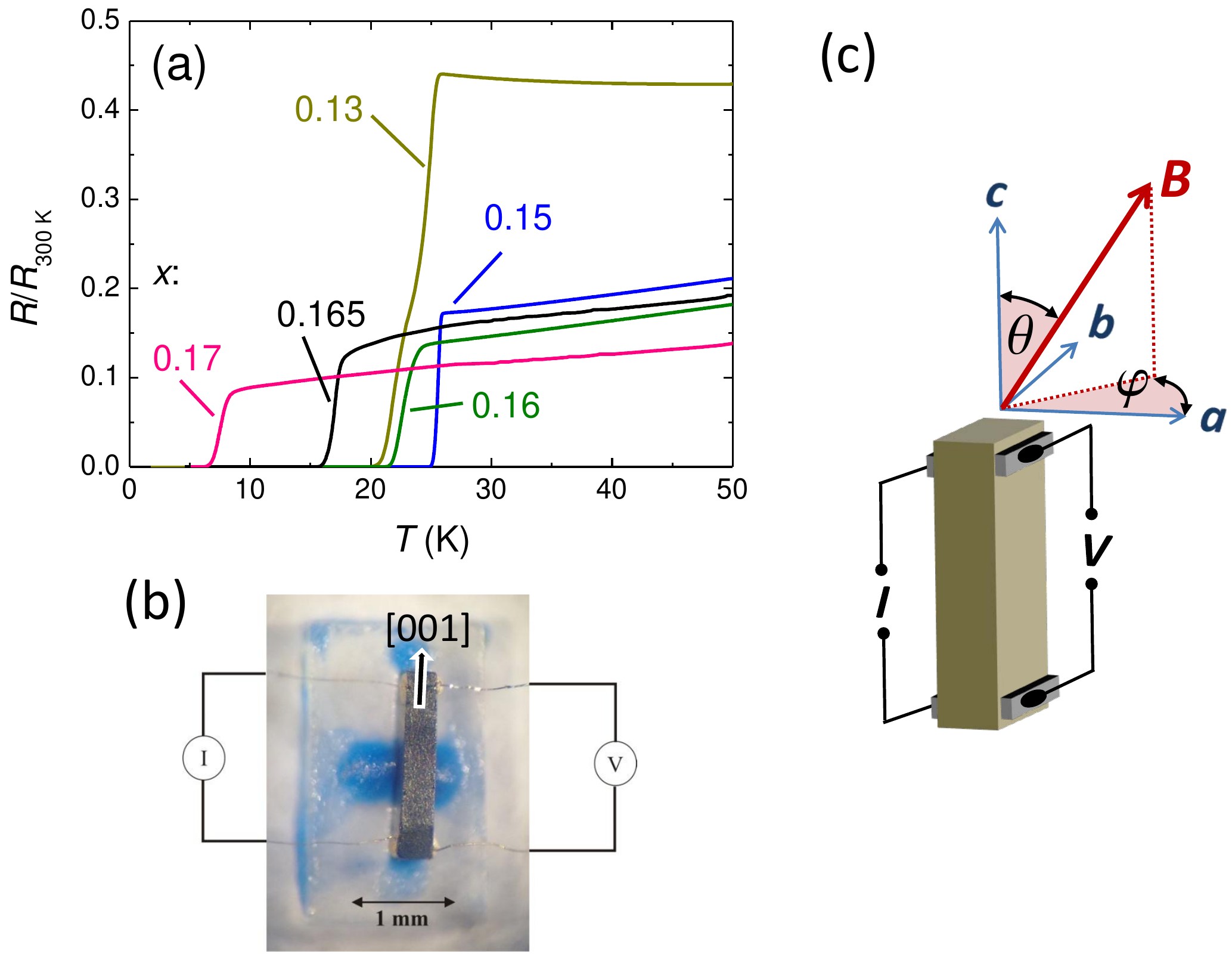}
\caption{
(a) Temperature-dependent interlayer resistance, normalized to its room-temperature
value, for typical Nd$_{2-x}$Ce$_x$CuO$_{4}$ crystals
studied in this work. The labels 0.13 to 0.17 indicate the Ce concentration
$x$, which is considered to be equal to the electron doping level~\cite{lamb10}.
(b) Photograph of a NCCO crystal with current ($I$) and voltage ($V$) leads
mounted on a sapphire plate and prepared for high-field measurements. Note that
the largest dimension of the sample is along the $c$-axis. (c) Schematic view
of the experimental geometry. The current is applied parallel to the $c$-axis;
angles $\theta$ and $\varphi$ define the orientation of the magnetic field with
respect to the crystal axes. }
 \label{fig2}
\end{figure}

The resistance of the samples was measured out-of-plane, that is in the
direction perpendicular to the conducting CuO$_2$ layers, for two reasons.
First, for the angle-dependent magnetoresistance studies: the AMRO phenomenon
mentioned in Section~\ref{sec:intro} is an inherent property of the interlayer
magnetoresistance and should be much more pronounced in this
configuration~\cite{kart04,kuri92}. Second, due to the high resistivity
anisotropy, $\sim 10^3$, the interlayer resistance value is usually much higher than
the in-plane resistance and, hence, easier to measure. Taking the advantage of
the TSFZ technique{\footnote{The as-grown specimen is a long rod, typically
about 10\,cm long and 0.5\,cm in diameter and consists of just a few single
crystal grains~\cite{lamb10}.}}, we were able to further increase the signal by
optimizing the shape of the crystal cut out of the as-grown rod. Given the
room-temperature resistivity of NCCO in the range $3$ to
$6\,\Omega$\,cm, depending on the doping level, and the typical sample
dimensions, about $0.3\times 0.3\times 1\,\mathrm{mm}^3$ with the largest
dimension along the $c$-axis, we could obtain a resistance of
$300-600\,\Omega$ at room temperature and $30-100\,\Omega$ at low temperatures.
For some batches we had to decrease the sample thickness to avoid crystal
inhomogeneity. In those cases the low-temperature resistance values were $\sim
1-10\,\Omega$. These values are still quite convenient and can be measured
accurately.

Fig.~2(b) shows a NCCO crystal prepared for high-magnetic-field measurements.
For voltage and current leads, annealed $20\,\mu$m Pt wires were attached using
either Dupont 4929 silver paste or Epo-Tek H20E conducting epoxy. The latter
provided the most reliable contacts with the lowest resistance. Due to their
highly anisotropic susceptibility caused by the magnetic moments of the
Nd$^{3+}$ ions, NCCO crystals experience a very strong torque in a tilted
magnetic field. Therefore, for our experiments we firmly glued the samples to a
sapphire plate by Stycast 2860~FT (blue color), as shown in Fig.~2(b). The
plate was then fixed by the same epoxy to an appropriate sample holder.

The angle-dependent magnetoresistance was measured in steady fields up to 28
and 34~T provided by the 20~MW resistive magnet at the LNCMI-Grenoble.
The samples were mounted onto a home-made two-axes sample
rotator allowing an {\it in situ} rotation at a fixed $B$, at temperatures
between 50 and 1.4~K. The resistance was measured as a function of polar angle
$\theta$ between the field direction and the crystallographic $c$-axis for
different fixed azimuthal angles $\varphi$, as shown in Fig.~2(c). SdH
oscillations were studied with the magnetic field applied parallel to the
$c$-axis. These experiments were done in pulsed fields provided by the Dresden
and Toulouse high-field facilities as well as in steady fields at Grenoble.

\section{Angle-Dependent Semiclassical Magnetoresistance}
\label{sec:magres}

\begin{figure}[tb]
\centering
\includegraphics[width=0.9\columnwidth]{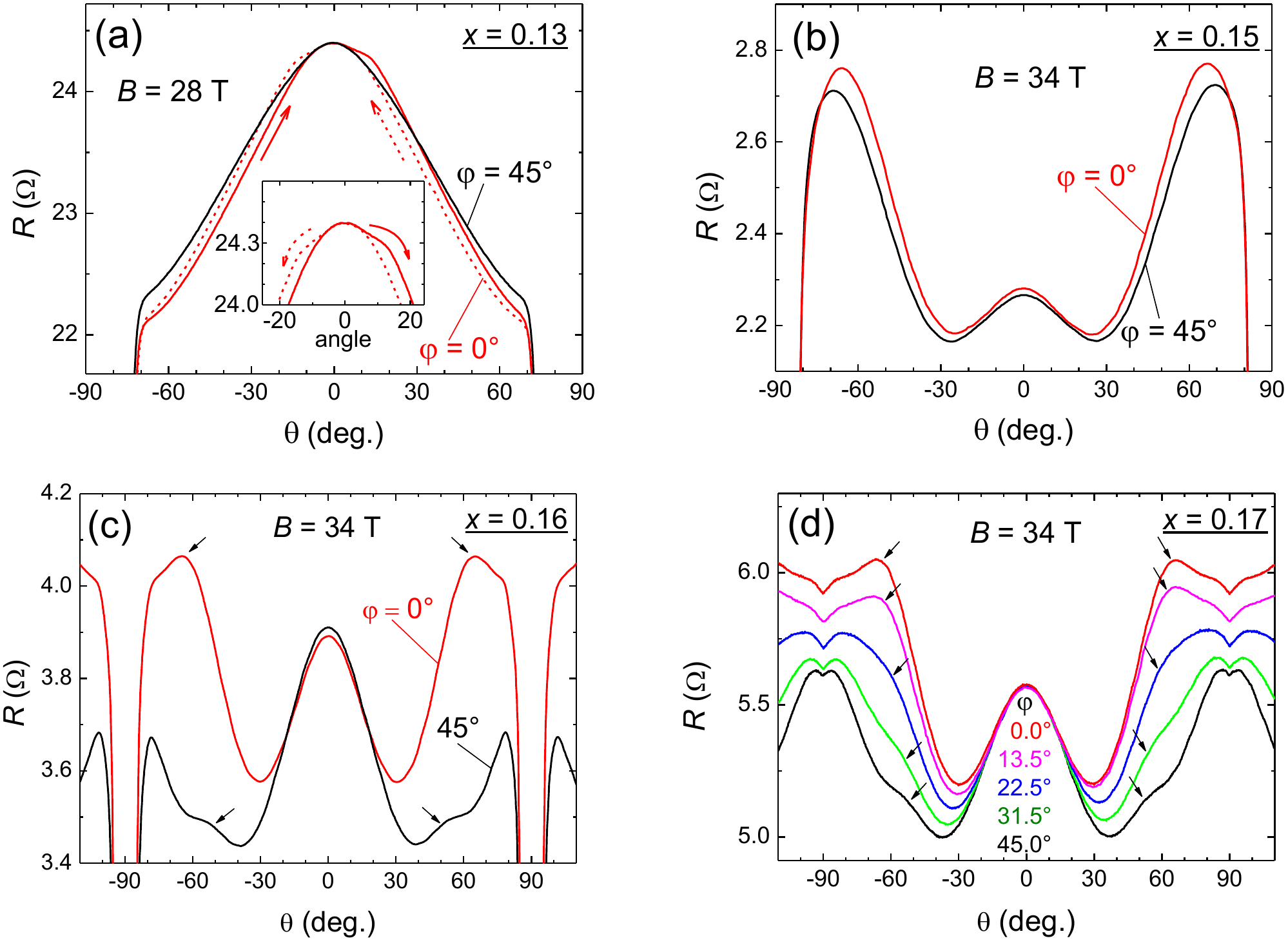}
\caption{
Angle-dependent interlayer magnetoresistance of
Nd$_{2-x}$Ce$_x$CuO$_{4}$ single crystals with doping level $x=0.13$
(a), 0.15 (b), 0.16 (c), and 0.17 (d) recorded at $T=1.5$\,K and applied
magnetic fields of 28\,T (a) and 34\,T (b) to (d). The overdoped samples (c)
and (d) exhibit AMRO features (marked by arrows) with their $\theta$ positions
being dependent on the azimuthal angle $\varphi$. }
 \label{fig3}
\end{figure}

Fig.~\ref{fig3} shows examples of the angle-dependent interlayer
magnetoresistance of NCCO for different doping levels, recorded at high
magnetic fields and $T=1.5$\,K. At field directions almost parallel to the
conducting CuO$_2$ plane (i.e. around $\theta=\pm 90^{\circ}$) the resistance
drops down due to the onset of superconductivity. Outside the superconducting
angular range, the $R(\theta)$ curves exhibit different shapes, depending on Ce
concentration. The typical behavior of underdoped samples is shown in
Fig.~3(a). The resistance is maximum for the field direction perpendicular to
the CuO$_2$ planes, i.e. for $\theta =0^{\circ}$, and then gradually decreases
at tilting the field towards the in-plane direction, until superconductivity
sets in at high tilt angles [$|\theta|>70^{\circ}$ for $x =0.13$, $T=1.5$~K,
and $B=28$~T, as shown in Fig.~\ref{fig3}(a)]. The dome-like shape of the
$R(\theta)$ curve centered at $\theta =0^{\circ}$ is opposite to what one expects
for a typical layered metal. There, the magnetoresistance generally increases
with increasing $|\theta|$~\cite{kart04}. The anomalous $\theta$-dependence
obtained for NCCO is almost insensitive to the azimuthal orientation of the
applied field, that is, to the angle $\varphi$. This suggests that it is
determined solely by the out-of-plane field component. Such behavior is similar
to what was observed in some organic layered metals and associated with
incoherent interlayer charge transfer (for a recent discussion,
see~\cite{kart09}). In addition, the $R(\theta)$ dependence taken at $\varphi =
0^{\circ}$ exhibits a clear hysteresis with respect to the rotation direction
at $|\theta|\geq 6^{\circ}$. This hysteretic behavior is most likely related to
a field-induced reorientation
of ordered spins. Indeed, recent reports on the in-plane field
rotation~\cite{yu07,wu08} point to a significant role of spin ordering in the
magnetoresistance of underdoped NCCO. Notably, in our experiments signatures of
spin ordering are present even in the superconducting compound with $x=0.13$.
Further work is necessary for clarifying the detailed mechanism responsible for
coupling the interlayer magnetoresistance to electron spins.

\begin{figure}[tb]
\centering
\includegraphics[width=0.8\columnwidth]{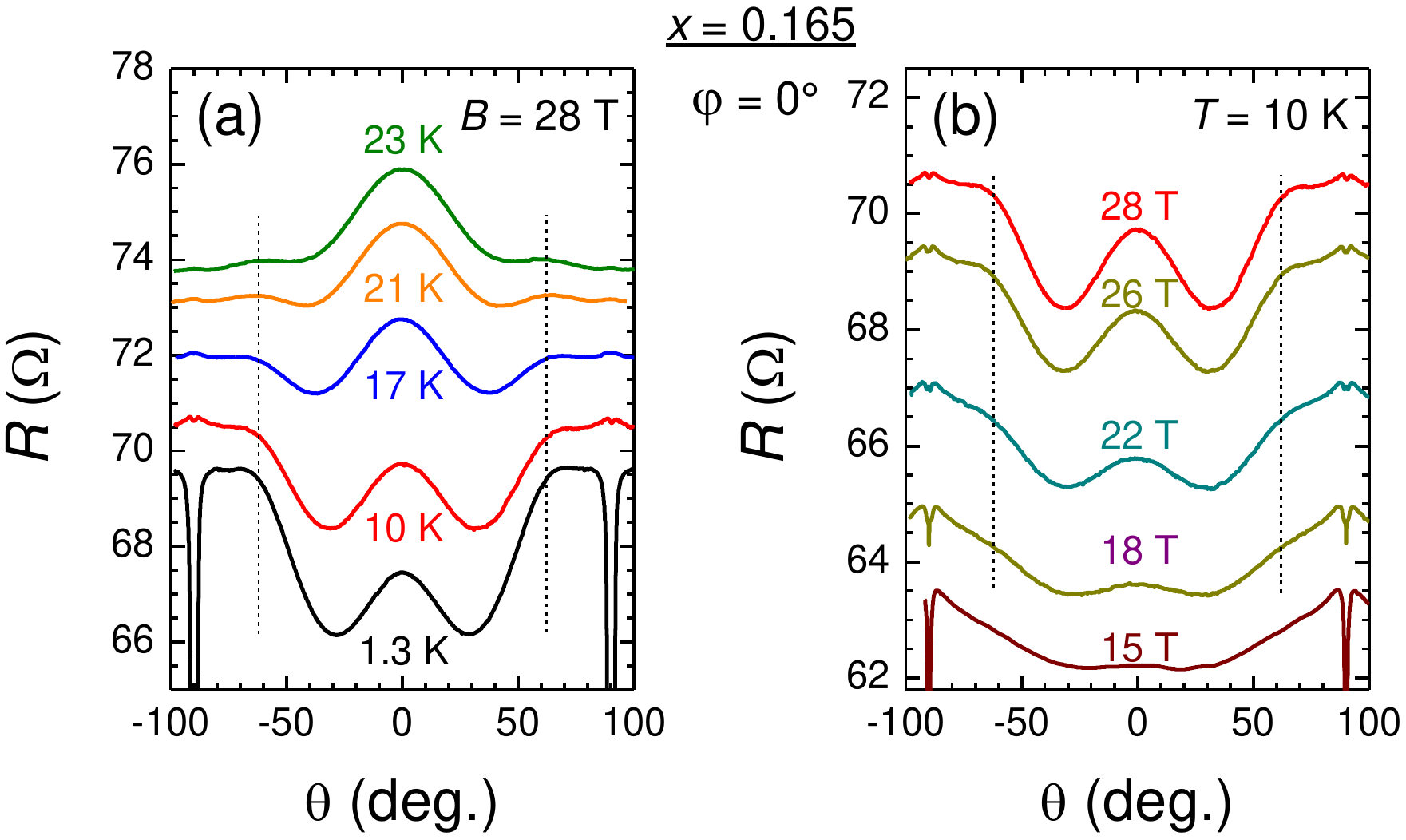}
\caption{
Angle-dependent magnetoresistance of a Nd$_{2-x}$Ce$_x$CuO$_{4}$ single
crystal with $x=0.165$ taken at different temperatures (a) and different
magnetic field strengths (b). One can see that the AMRO positions (marked by
vertical dashed lines) are independent of $T$ and $B$. }
 \label{fig4}
\end{figure}

As we increase the Ce concentration to the optimal and, further on, to the
overdoped regime, the anomalous contribution to the magnetoresistance weakens,
giving way to the conventional behavior associated with the orbital effect of
the applied magnetic field on the charge carriers. This causes a positive,
$\varphi$-dependent slope $dR/d|\theta|>0$ of the angular dependence
$R(\theta)$ over an extended angular range $30^{\circ}\lesssim|\theta|\lesssim
80^{\circ}$, as shown in Fig.~\ref{fig3}(b) to (d) for samples with
$x=0.15$, $0.16$, and $0.17$. The most interesting feature in the overdoped
regime is a shallow hump (marked by arrows) superposed on the monotonic slope at
$|\theta| \simeq 53-62^{\circ}$, depending on $\varphi$. The sensitivity of the
angular position of the hump to the azimuthal orientation and its independence
of temperature and the magnetic field strength, as is shown in Fig.~\ref{fig4},
clearly point to the geometrical origin of this feature.
Therefore, we attribute this hump-like feature to the AMRO effect.

The $R(\theta)$ curves in Fig.~\ref{fig3}(c),(d) are qualitatively similar
to the angle-dependent magnetoresistance of the hole-overdoped cuprate
Tl2201 \cite{huss03,huss06}, which also exhibits a local maximum around
$\theta = 0^{\circ}$ and a shoulder at $|\theta| \simeq 30-40^{\circ}$.
In Tl2201 having the same body-centered tetragonal crystal symmetry,
the entire angular dependence can be nicely described within
the semiclassical Boltzmann transport model with a $\mathbf{k}$-dependent
scattering time \cite{huss06,anal07}. In particular, both the central hump
and the side feature come from the AMRO effect of the slightly warped large
Fermi cylinder centered at the corner of the Brillouin zone. By analogy,
it is tempting to consider our data on NCCO as a manifestation of the
unreconstructed large Fermi cylinder, as in Fig.~\ref{fig1}(b). However,
the situation seems to be more intricate. By contrast to the case of
Tl2201, the central ($\theta=0^{\circ}$) hump is much more pronounced,
as compared to the side features. Furthermore, at odds with what is
expected from an AMRO peak, it does not diminish at increasing the
temperature. Instead, it develops in size and becomes a dominating
feature in the angular dependence above $\sim 20$\,K, as shown, e.g.,
in Fig.~\ref{fig4}(a) for a sample with $x=0.165$. The 21 and 23\,K
curves in this Figure rather resemble the $R(\theta)$ dependence in the
underdoped, $x=0.13$ NCCO, c.f. Fig.~\ref{fig3}(a). It is, therefore,
likely \cite{helm10-} that, at least at elevated temperatures, the
central hump originates from a remnant of the anomalous, presumably
incoherent, interlayer transport channel rather than from the conventional
AMRO effect.

\begin{figure}[tb]
\centering
\includegraphics[width=0.6\columnwidth]{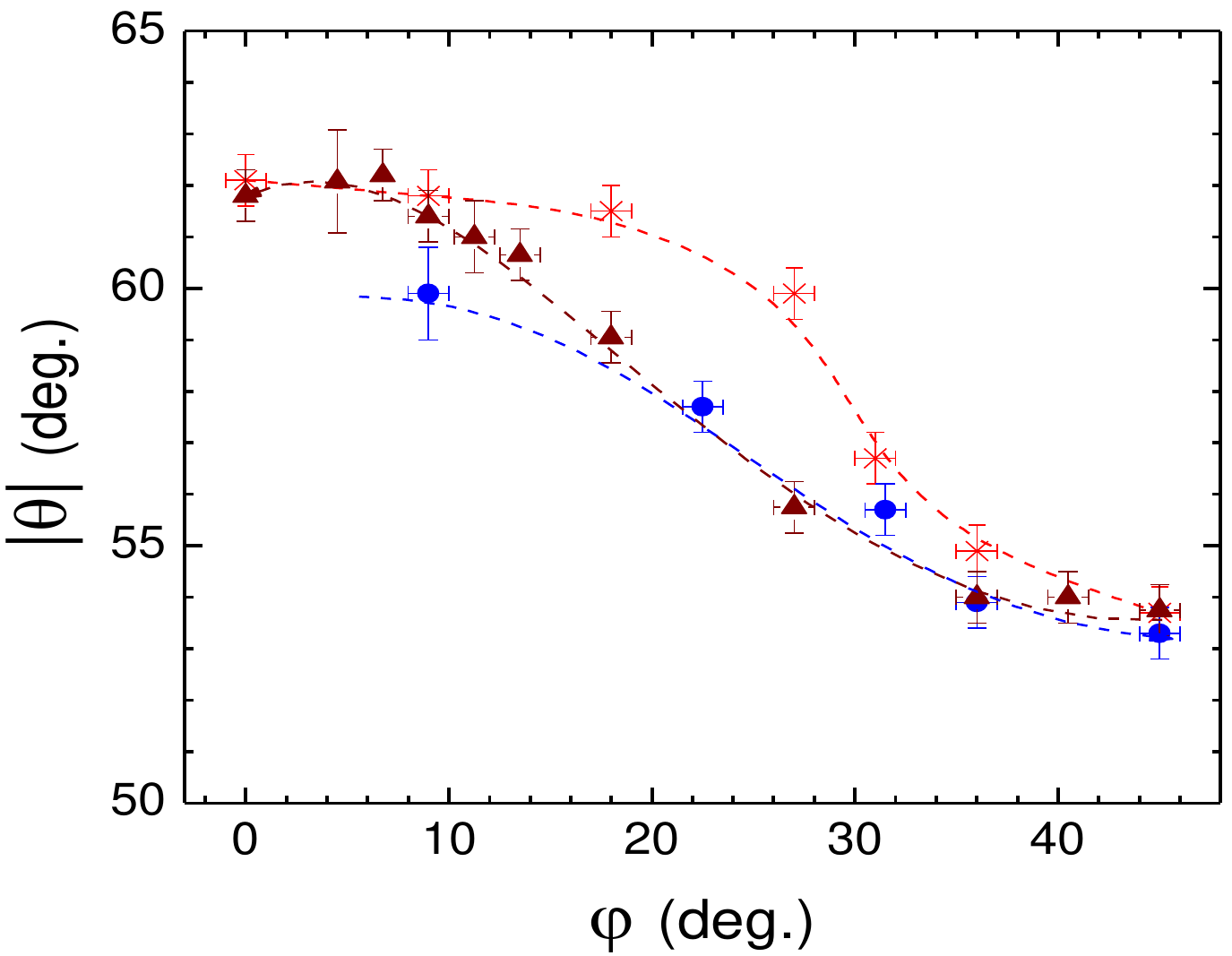}
\caption{
Angular position $|\theta|$ of the AMRO features in the $R(\theta)$ curves
plotted versus the azimuthal angle $\varphi$ used in the $R(\theta)$ sweeps for
Nd$_{2-x}$Ce$_x$CuO$_{4}$ single crystals with doping levels $x=0.16$
(blue circles), 0.165 (brown triangles), and 0.17 (red stars). The dashed lines
are guides to the eye. }
 \label{fig5}
\end{figure}

The superposition of the conventional orbital and anomalous contributions
to the $R(\theta)$ dependence makes it problematic for a quantitative
analysis. However, already at the present stage,
we can assert that the Fermi surface giving rise to the AMRO is the same
in the doping interval from $x=0.16$ to 0.17. This is evidenced by the
similarity of the AMRO features in Fig.~\ref{fig3}(c),(d) and Fig.~\ref{fig4},
as well as by their angular positions shown in Fig.~\ref{fig5}. Having in mind
the strong difference between the quantum-oscillation spectra reported for
$x=0.16$ and 0.17~\cite{helm09}, one is lead to the conclusion that the large
cyclotron orbit, indistinguishable from that on the original Fermi cylinder
shown in Fig.~\ref{fig1}(b), and the small orbit corresponding to hole pockets
of the reconstructed Fermi surface [see Fig.~\ref{fig1}(c)] should coexist at
one and the same composition. Indeed, this was confirmed by the observation of
both SdH frequencies in the strongly overdoped ($x=0.17$) NCCO crystal~\cite{helm10-}.
In Section~\ref{sec:sdh}, we present further data on quantum oscillations which
provides clear evidence for magnetic breakdown to take place also at lower
doping down to $x=0.16$.

We note that for $x=0.15$ the $R(\theta)$ dependence presented in Fig.~3(b)
does not show clear AMRO features up to $B=34$\,T. This seems to favor the
scenario of the large magnetic breakdown orbit being responsible for the AMRO:
with decreasing $x$ from 0.16 to 0.15, the superlattice potential is enhanced,
leading to an increase of the energy gap between different parts of the
reconstructed Fermi surface and thereby impeding magnetic breakdown.
The suppression of magnetic breakdown could be a reason for the absence of
the AMRO at $x=0.15$, if the latter were caused by the breakdown orbit.
On the other hand, one should keep in mind that the composition $x=0.15$ is closer
to the underdoped regime, in which the magnetoresistance is dominated
by the anomalous incoherent mechanism. This should also suppress the AMRO
effect even if it is caused by small orbits on the reconstructed Fermi surface.
In this context, it would be interesting to study the angle-dependent
magnetoresistance of the $x=0.15$ sample at even higher fields up to 45\,T.

\section{Shubnikov-de Haas (SdH) Oscillations}
\label{sec:sdh}

\begin{figure}[tb]
\centering
\includegraphics[width=0.9\columnwidth]{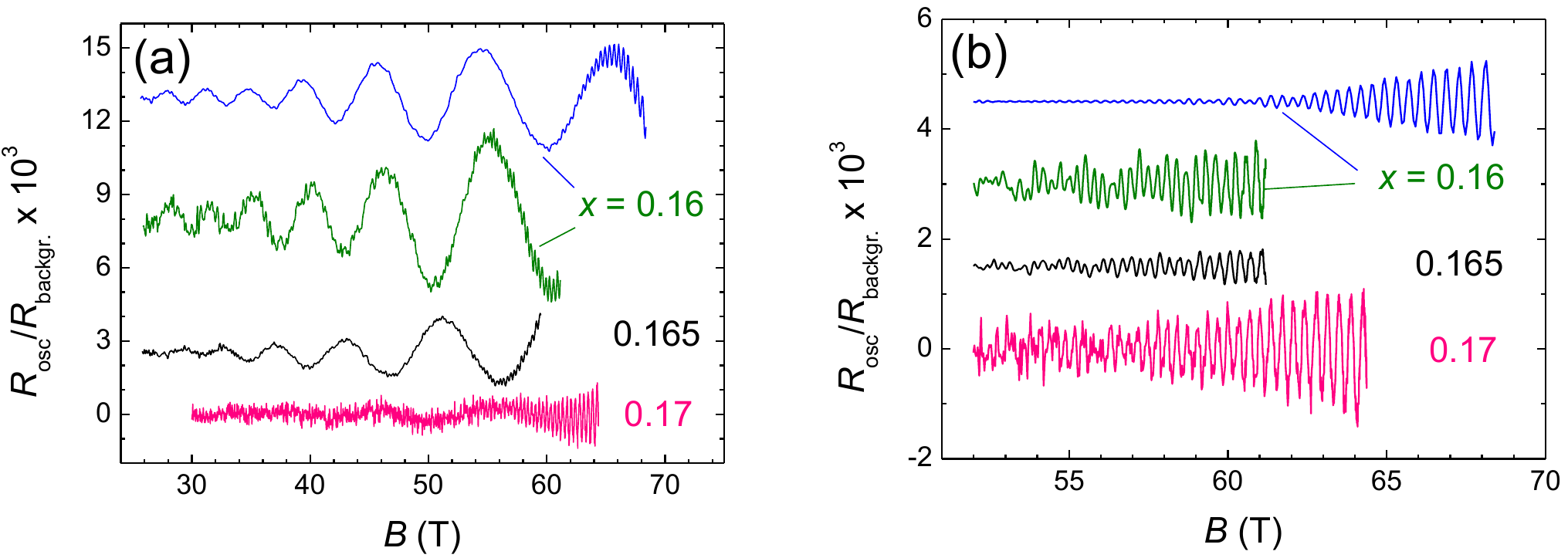}
\caption{
(a) Oscillatory component of the interlayer magnetoresistance at pulsed fields
applied perpendicular to the CuO$_2$ layers. The data was taken at $T=2.5$\,K for
Nd$_{2-x}$Ce$_x$CuO$_{4}$ single crystals with doping levels $x=0.16$,
0.165, and 0.17, and for another sample with $x=0.16$ at the Dresden and at the Toulouse (blue curve)
high-field facilities, respectively. In addition to slow SdH oscillations, one can resolve fast
oscillations emerging from $B\simeq 55$\,T for all three doping levels. (b) Fast SdH oscillations
extracted from the data in (a) at $B>54$\,T by filtering out the low-frequency component.
The data in (a) and (b) are vertically shifted for better visibility}
 \label{fig6}
\end{figure}

\begin{figure}[tb]
\centering
\includegraphics[width=0.6\columnwidth]{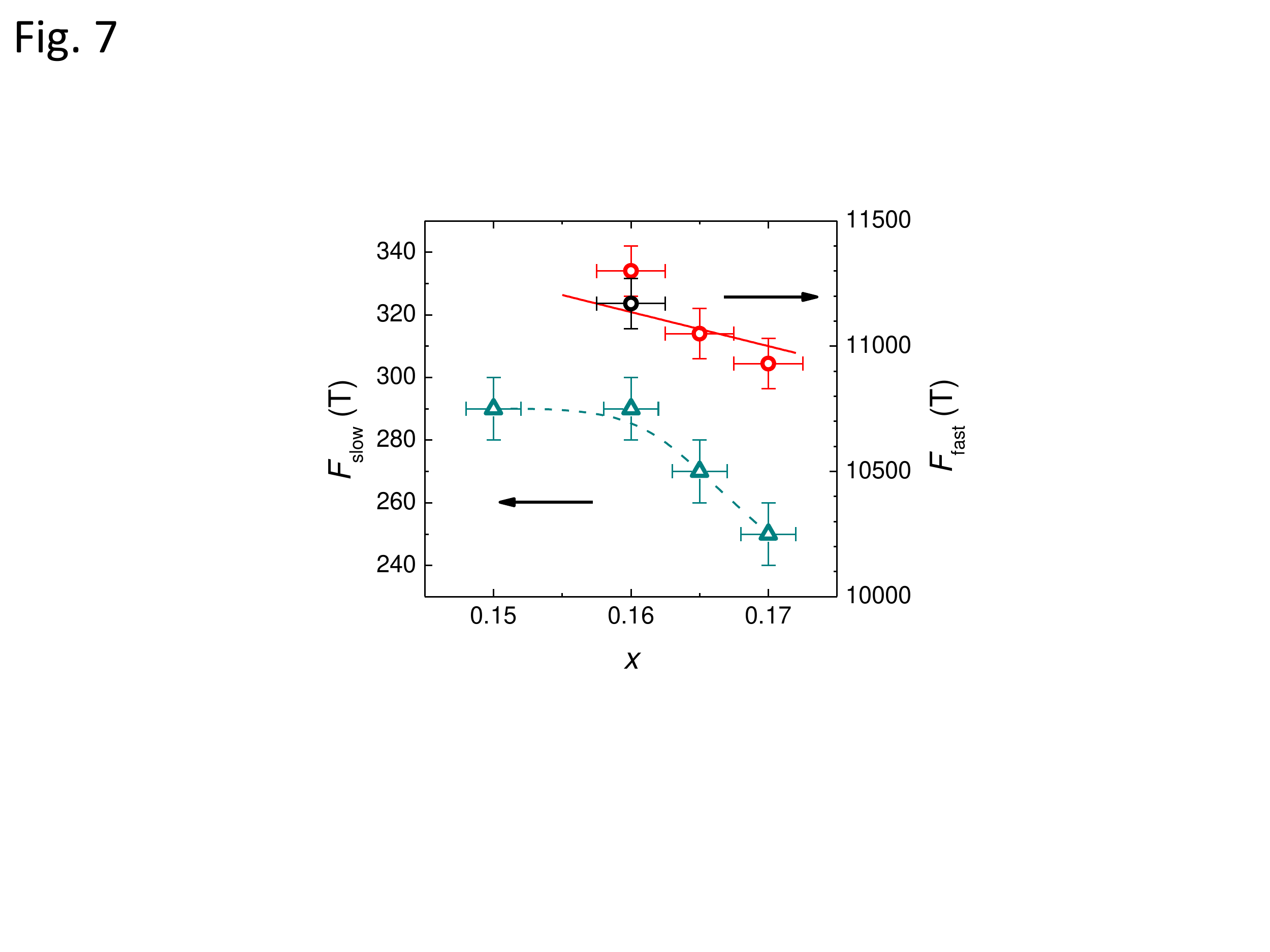}
\caption{
Dependence of the frequency $F$ of the slow (triangles, left-hand scale) and
fast (circles, right-hand scale) oscillations on the doping level $x$. The
$F_{\rm slow}$ data points summarize results obtained on several
Nd$_{2-x}$Ce$_x$CuO$_{4}$ single crystals of each doping level in pulsed
and steady fields. The dashed line is a guide to the eye. The $F_{\rm fast}(x)$
data points have been obtained in pulse-field experiments at the Dresden (red
circles) and Toulouse (black circle) high-field facilities. The solid line
shows the $F(x)$ dependence calculated from the nominal doping level according
to the large Fermi surface shown in Fig.~\ref{fig1}(b). }
 \label{fig7}
\end{figure}

Motivated by the observation of AMRO suggesting magnetic breakdown in
overdoped NCCO, we performed measurements in pulsed magnetic fields to search
for multiple SdH frequencies. Fig.~\ref{fig6}(a) shows the oscillating
component of the interlayer magnetoresistance (normalized to the field-dependent
background) for $x=0.16$, 0.165, and 0.17. All the curves exhibit oscillations
with two different frequencies. The slow oscillations, similar to those reported
earlier for $x=0.15$ and 0.16~\cite{helm09}, are associated with the small hole
pockets of the reconstructed Fermi surface [see Fig.~\ref{fig1}(c)]. The
fast oscillations, shown on an enlarged scale in Fig.~\ref{fig6}(b) have
frequencies of about 11\,kT. 
This is obviously
consistent with the cross-sectional area of the large unreconstructed Fermi
cylinder. The oscillation frequencies clearly depend on $x$. This dependence,
plotted in Fig.~\ref{fig7} for both the slow and fast oscillations,
rules out the possibility that the two frequencies are caused by a coexistence
of two phases with different Fermi surfaces. Indeed, imagine a sample
consisting of two phases with different carrier concentrations giving
rise to different Fermi surface topologies. Then, a change of the nominal
doping level $x$ would lead to changing the volume fractions of the phases
and, thus, affect the amplitudes of the SdH oscillations but not their
frequencies. This scenario obviously contradicts the regular $x$ dependence
of the oscillation frequencies shown in Fig.~\ref{fig7}. Moreover, within our experimental
resolution, the high frequency coincides with what is expected from the Luttinger
rule for the given doping levels (solid line in Fig.~\ref{fig7}).
Thus, we conclude that both SdH frequencies are an inherent feature of each
doping level in the range $x = 0.16-0.17$, rather than a result of a phase mixture.
An important implication of this conclusion is that the Fermi surface remains
reconstructed up to the highest doping level, $x=0.17$, and the fast
oscillations originate from magnetic breakdown through small gaps between the
electron and hole pockets.

\begin{figure}[tb]
\centering
\includegraphics[width=1.0\columnwidth]{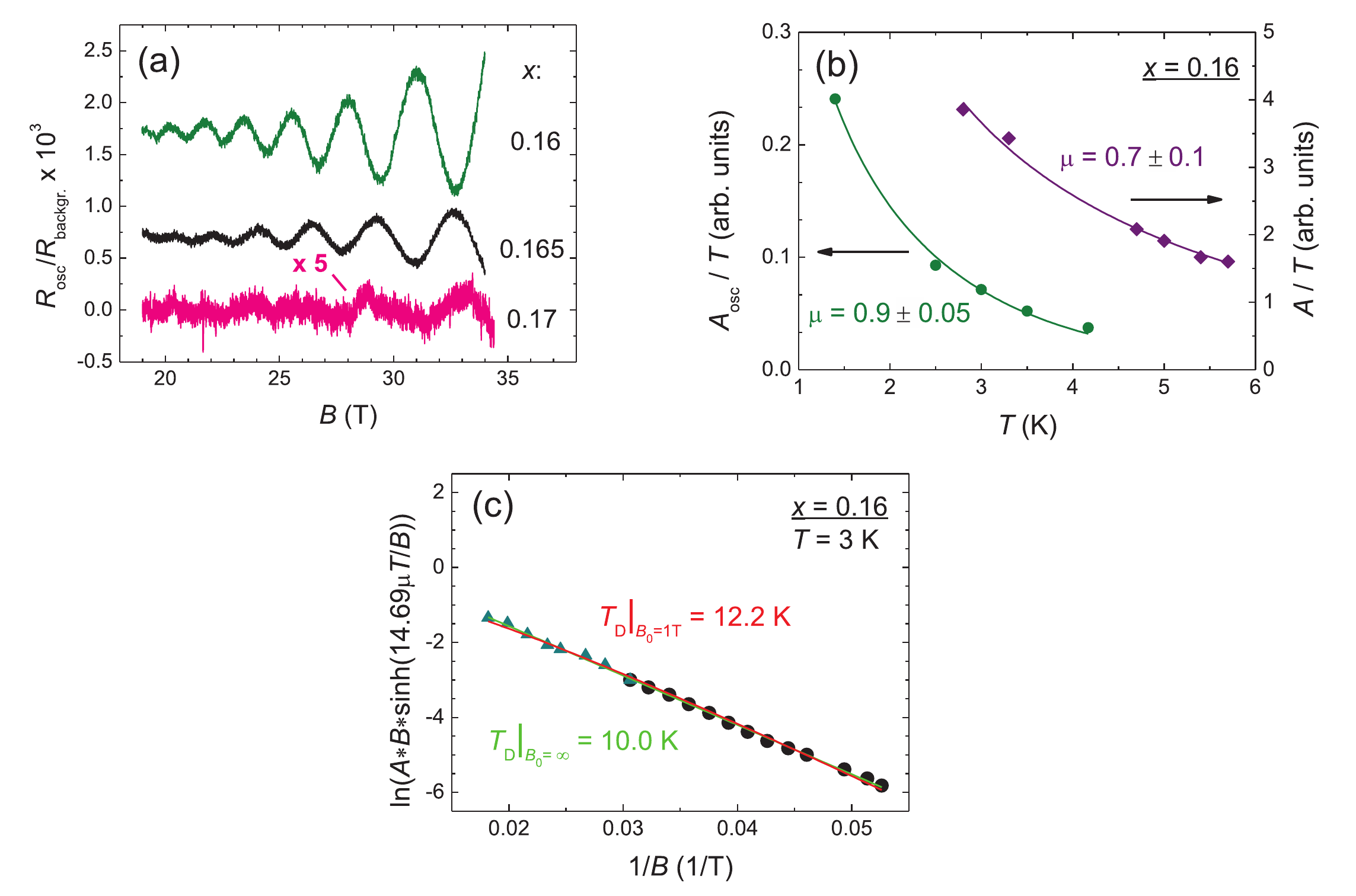}
\caption{
(a) Slow SdH oscillations in overdoped Nd$_{2-x}$Ce$_x$CuO$_{4}$ single
crystals recorded in steady magnetic fields at $T=1.5$\,K. The data are vertically
shifted for better visibility.
(b) Temperature dependence of the oscillation amplitude $A_{\rm osc}$ obtained from the
steady-field (circles) and pulsed-field (diamonds) experiments. Solid lines are
Lifshitz-Kosevich fits yielding the effective cyclotron mass values $\mu
\approx 0.9$ and 0.7 for the steady- and pulsed-field data, respectively, see
text. (c) Dingle plot for the slow-oscillation amplitude including pulsed-
(triangles) and steady-(circles) field data. The fits to the theoretical field
dependence assuming a magnetic breakdown field $B_0 =\infty$ (green line) and
1\,T (red line) are almost indistinguishable and give just slightly different values
for the Dingle temperature, see text.}
 \label{fig8}
\end{figure}

The SdH oscillations shown in Fig.~\ref{fig6} are very weak, $\sim
10^{-3}$ of the background resistance. To achieve the necessary resolution, we
had to apply rather high currents of up to 10\,mA. This caused considerable overheating which
was estimated by recording $R(B)$ curves at low-field pulses for
different currents and comparing them with steady-field data. On the
other hand, one can see from Fig.~\ref{fig6}(a) that the field dependence of
the slow oscillations is not very steep: they can still be well resolved, at
least for $x=0.16$ and 0.165 at $B\sim 30-35$\,T. Therefore, we have performed
a SdH experiment in steady fields up to 35\,T, which typically provide a lower noise level.
Fig.~\ref{fig8}(a) shows SdH oscillations for
three doping levels recorded at steady fields with a transport current of
0.4\,mA. This current was proved to cause no significant overheating effect
($\Delta T<0.1$ K). 

The oscillations frequencies, $F_{\mathrm{slow}}=290$, 270, and 250\,T for 
$x=0.16$, 0.165, and 0.17, respectively, perfectly coincide with the 
pulsed field data. For $x=0.17$, the oscillation amplitude barely exceeds the
noise level even at the highest fields. However, for the other two compositions
the slow oscillations can be observed down to below 20\,T and are suitable for
quantitative analysis.

By contrast to hole-underdoped YBa$_2$Cu$_3$O$_{6.5\pm\delta}$, for which
several different low SdH frequencies have been
reported~\cite{seba10,rams10,audo09}, the present NCCO compound only shows a
single frequency. In particular, so far we have not found any signature of
oscillation beating, caused by a slight warping of the Fermi surface in the
interlayer direction. This suggests that the distance between adjacent Landau
levels, $\hbar \omega_c$ (where $\omega_c=eB/m_c$ is the cyclotron frequency
and $m_c$ is the relevant effective cyclotron mass), exceeds
the interlayer dispersion at fields $B\gtrsim 20$\,T. Therefore, one should
take into account the strong two-dimensionality of the system in the analysis
of the oscillations. Generally speaking, the quasi-two-dimensional SdH effect
is a very complex problem and has not yet received a comprehensive
description even within the standard Fermi liquid theory~\cite{kart04,grig03}.
Moreover, it was recently pointed out \cite{thom10} that the presence of
strong electron correlations should dramatically affect the behavior of
quantum oscillations, provided $\hbar \omega_c$ significantly exceeds the
interlayer transfer integral $t_{\perp}$ and scattering-induced broadening of
Landau levels $\sim\hbar/\tau$ (where $\tau$ is the scattering time).
However, our case is simplified due to the weakness of the oscillation amplitude
and the absence of higher harmonics, which is indicative of a strong Landau level 
broadening. 
Therefore, we adopt the so-called two-dimensional Lifshitz-Kosevich (LK)
formula~\cite{shoe84} for quantum oscillations in a two-dimensional metal
with a constant chemical potential. It should be valid for the in-plane SdH effect,
and we assume both the oscillating and monotonic components of the interlayer
conductivity to be proportional to those of the in-plane conductivity. Then
the field and temperature dependence of the oscillation amplitude are expressed
as~\cite{shoe84} $A_{\rm osc} \propto R_{\rm T}R_{\rm D}R_{\rm MB}$, where
$R_{\rm T}$ and $R_{\rm D}$ are the standard temperature and Dingle reduction
factors, respectively, and $R_{\rm MB}$ describes the influence of magnetic
breakdown.\footnote{We do not consider the spin reduction factor $R_{\rm
S}$~\cite{shoe84}. It mainly affects the angle dependence of the oscillation amplitude
which is not studied here.}

From the $T$-dependence of the oscillation amplitude, given by $R_{\rm T}=(K\mu
T/B)/\sinh(K\mu T/B)$ (where $K=2\pi^2km_e/\hbar e\approx 14.69\,{\rm T/K}$,
$\mu=m_c/m_e$ is the effective cyclotron mass normalized to the free electron
mass $m_e$, $k$ is the Boltzmann constant, and $e$ is the elementary charge),
 one can estimate the effective cyclotron mass.
Fig.~\ref{fig8}(b) shows the experimental data for $x=0.16$, fitted to
the LK formula. From the steady-field data (circles) we derive $\mu_{0.16} =
0.92\pm 0.05$. A similar value, $\mu_{0.165} = 0.9\pm 0.05$, is obtained for
$x=0.165$. For comparison, the effective mass plot obtained from the pulsed-field
data is shown by diamonds in Fig.~\ref{fig8}(b), giving a somewhat lower
value, $\mu_{0.16}^{\rm pulse} = 0.7 \pm 0.1$. The discrepancy is apparently
caused by an uncertainty in the sample temperature in the pulse-field
experiment. Although the induced eddy currents do not cause a significant
heating effect due to the relatively long pulse duration, $\simeq 0.2$\,s, the 
temperature is affected by the fast orientation of the paramagnetic Nd$^{3+}$
ions reducing the entropy of the spin system. Obviously, the effect is enhanced
at lower temperatures, which results in a flatter apparent $T$-dependence and,
hence, in a smaller $\mu$. Therefore, the derivation of the cyclotron mass from
the steady-field data is considered more reliable. Unfortunately, the fast
oscillations have so far been detected only in pulsed fields above 50\,T. The
estimate of the cyclotron mass from this pulse-field data yields $\mu = 2.2 \pm
0.2$ for $x=0.17$ and $2.4\pm0.2$ for $x=0.165$. We should, however, keep in
mind that the real values may be about $20$\% higher.

After having derived the cyclotron mass, we can use the field dependence of the
oscillation amplitude to estimate the Dingle temperature, $T_{\rm D}$, determined
by the scattering-induced broadening of Landau levels. The corresponding Dingle
plot for $x=0.16$ is shown in Fig.~\ref{fig8}(c), yielding $T_{\rm D} \approx
10$\,K. Strictly speaking, when fitting the $B$-dependent SdH amplitude we
should take into account the effect of magnetic breakdown through the gap (which
is thus considered as a breakdown junction) between the hole and electron pockets
of the reconstructed Brillouin zone. The closed orbit on the small hole pocket
responsible for the slow oscillations involves reflections from two breakdown
junctions. Therefore, the corresponding damping factor for the oscillation
amplitude is \cite{shoe84,fali66}: $R_{\rm MB}^{\rm slow} = [1-\exp(-B_0/B)]$,
where $B_0$ is the characteristic breakdown field. Including $R_{\rm MB}$
with a very low breakdown field, $B_0=1$\,T, (see below) into the field dependence
$A_{\rm osc}(B)$, we obtain the fit shown by the red line in
Fig.~\ref{fig8}(c), which is almost indistinguishable from the pure
[$B_0=\infty$, green line in Fig.~\ref{fig8}(c)] Dingle fit and gives just a
slightly different $T_{\rm D} \approx 12$\,K. Using the conventional
relationship between $T_{\rm D}$ and the scattering time $\tau$, $T_{\rm
D}=\hbar/2\pi k_{\rm B}\tau$, we estimate $\tau \approx 0.15$\,ps. This
corresponds to a mean-free path  averaged over the cyclotron orbit of
$\langle\ell\rangle \simeq 18$\,nm.

The obtained cyclotron mass and scattering time values determine
the field strength parameter $\omega_c\tau$ which merely reaches unity at
$B \simeq 35$\,T (for the slow oscillations; for the fast SdH oscillations
it is even lower due to heavier $m_c$). This justifies the validity of the
standard two-dimensional LK formula for fitting the steady field data. However,
as mentioned above, one can expect deviations from the LK theory at higher
fields, when $\omega_c\tau > 1$. In particular, in future it would be
interesting to discuss the pulsed field data in terms of the recent theory~\cite{thom10}
taking into account strong electron interactions.

As follows from the comparison between the two fits in Fig.~\ref{fig8}(c),
it is very difficult to extract the breakdown field from
the  $B$-dependence of the slow oscillation amplitude
in the given restricted field range. For the fast
oscillations, the breakdown reduction factor is determined by tunneling
through 8 equivalent magnetic breakdown junctions and has the form
\cite{shoe84,fali66}:
$R_{\rm MB}^{\rm fast} = \exp(-4B_0/B)$. One can see that it has exactly
the same functional field dependence as the Dingle factor, $R_{\rm
D} = \exp(-K\mu T_{\rm D}/B)$, which makes it impossible to evaluate $T_{\rm
D}$ and $B_0$ separately. Nevertheless, we can make a rough estimate by
simply comparing the amplitudes of the fast and slow oscillations at a given
field, say, $B=60$\,T. Assuming the same Dingle temperature for the carriers on
the small and large orbits and using the cyclotron mass values $\mu_{\rm slow}
= 0.9$ and $\mu_{\rm fast} = 2.2$, we can reproduce the ratio between the
oscillation amplitudes, $A_{\rm slow} / A_{\rm fast} \simeq 10$ observed for
$x=0.16$ and 0.165 (see Fig.~\ref{fig6}) by taking $B_0=7$~T. For the highest
doping level, $x=0.17$, the slow and fast oscillations have approximately equal
amplitudes at 60\,T, yielding $B_0 \simeq 1$\,T.
This estimate demonstrates that observation of both the high and low
frequencies in the SdH spectrum is only possible at $B\gg B_0$, i.e. in the strong
magnetic-breakdown limit. This is because the strong difference in the Dingle
factors for the slow and fast oscillations, originating from different cyclotron
masses and the large Dingle temperature, must be compensated by the difference in
the magnetic-breakdown reduction factors. In particular, for $x=0.17$ the ratio between
the breakdown factors at 60 T is:
$R_{\rm MB}^{\rm fast}/R_{\rm MB}^{\rm slow} \simeq 60$.

Here, it is worth making a remark concerning the contribution of the electron
pockets of the reconstructed Fermi surface [see Fig. 1(c)] to the SdH effect. In the
strong magnetic breakdown regime, $B\gg B_0$, the probability of encircling
a full closed orbit around an electron pocket, involving reflections at 4 breakdown
junctions, is strongly suppressed. Taking the above estimate,
$B_0 = 7$\,T for $x=0.16$, we arrive at the breakdown reduction factor
$R_{\rm MB}^{\rm el.\,pocket}=\left[1-\exp(-B_0/B)\right]^2 \approx 10^{-2}$
at $B=60$\,T, which is  $\simeq 10$ times smaller than for the oscillations from
the small hole pockets at the same field (we recall that the latter require only
two reflections from magnetic breakdown junctions). Additional damping comes from the
temperature and Dingle factors, since the cyclotron mass corresponding to the
electron orbits is, obviously, heavier than that of the small hole orbits.
Altogether, we expect the amplitude of the oscillations coming from the
electron pockets to be about two orders of magnitude lower than that
associated with the hole pockets at $x=0.16$ and even further reduced at
higher doping levels. It is, therefore, not surprising that no clear signature
of the electron pockets has been found in the SdH spectrum so far.

Using the Blount criterion, one can estimate the energy gap $\Delta$ between the
electron and hole bands formed by the superlattice potential~\cite{blou62}:
$\Delta \simeq \left(\hbar e B_0\varepsilon_{\rm F}/m_{\rm c}\right)^{1/2}$, where
$\varepsilon_{\rm F}\approx 0.5$\,eV is the Fermi energy and
$m_{\rm c}\approx 2.2 m_{\rm e}$ is the cyclotron mass corresponding to the
large magnetic-breakdown orbit. With the above values $B_0$, one obtains
$\Delta \simeq 14$\,meV for $x = 0.16$ and 5\,meV for $x=0.17$.
While this is, of course, only a rough estimate, it shows that the superlattice
energy gap is very small, namely in the meV range.

\section{Concluding Remarks and Outlook}

The experimental results presented here for high-quality overdoped
Nd$_{2-x}$Ce$_x$CuO$_{4}$ single crystals provide convincing evidence
for the presence of magnetic breakdown. This indicates that the Fermi
surface is reconstructed up to the highest doping level ($x=0.17$) reliably
attainable for bulk NCCO. Note that this doping level corresponds to the very
end of the superconducting range in the phase diagram: $T_{\rm c}$ extrapolates
to zero at $x\approx 0.175$~\cite{lamb10}. Furthermore, our data show that the
superlattice potential is strongly reduced at the highest doping level.
Therefore, it is likely that both the superlattice potential and
superconductivity vanish at the same carrier concentration. This would strongly
suggest a close relation between the two ordering phenomena.

The existence of the superlattice potential in the overdoped regime
distinguishes our electron-doped compound from the hole-doped cuprates. In the
latter, a broken symmetry has been observed only below optimal doping so far.
The mechanism responsible for the broken translational symmetry is still to be
clarified. One possibility is that the commensurate antiferromagnetic ordering
persists from the underdoped regime~\cite{das08,sach10}. It was very
recently proposed~\cite{webe10} that, by contrast to hole-doped cuprates, the
mechanism underlying the metal-insulator transition in the electron-doped
materials is of magnetic origin rather than due to electronic correlations. In
this case the metallic and superconducting state could coexist with
antiferromagnetism, provided the magnetic interaction is not strong enough for
causing a metal-insulator transition. However, until now no static magnetic
ordering has been detected in overdoped NCCO~\cite{moto07}. This seems to favor
another scenario associated with a "hidden" $d$-density-wave
ordering~\cite{chak01,jia09}. On the other hand, it is possible that the ordering
only exists at high magnetic fields, i.e. at the conditions of our present
experiment. An in-depth study of the magnetic properties at high fields should
elucidate this issue.

Although our experiments already give new and interesting insight, further detailed
Fermi surface studies in high fields are highly important for clarifying
the origin of the superlattice potential. Angle-dependent
magnetoresistance experiments extending up to the highest steady fields of
45\,T and covering also a broader doping range, should allow a
quantitative characterization of the Fermi surface geometry and its dependence
on the carrier concentration. The angle dependence of the SdH oscillations is
also potentially important for understanding the spin state of the conduction
electrons and the nature of the ordering~\cite{seba10,rams10,garc10}.

As mentioned above, the suggested reconstruction of the Fermi surface in
overdoped NCCO implies the existence of electron pockets, in addition to the
hole ones. However, the magnetic breakdown occurring at the highest doping levels
suppresses orbits encircling the whole electron pocket more
effectively than those around hole pockets. Therefore, lower doping levels
with a stronger superlattice potential appear to be more appropriate for
searching for a manifestation of the electron pockets in high-field
magnetotransport.

Another still open question is how the Fermi surface develops upon
entering the underdoped range of the phase diagram. The lowest doping level at
which we have so far succeeded in finding SdH oscillations is $x=0.15$. The
reason why they are not observed at lower doping is not clear. Could it be that
the closed hole pockets disappear immediately below the optimal doping? Or is
it because the conventional orbital effect on the interlayer conductivity
becomes too small, compared to the emerging anomalous incoherent
magnetotransport? In the latter case, it might be more convenient to look for
magnetic quantum oscillations in the in-plane resistivity or in thermodynamic
properties. Unfortunately, the oscillations of magnetization (de Haas--van
Alphen effect) can hardly be detected in NCCO due to an overwhelming magnetic
contribution from the Nd$^{3+}$ moments. One could, however, search for the
oscillations in some other properties such as sound velocity or
magnetostriction.

\subsection*{Acknowledgments}

The work was supported by EuroMagNET II under the EU contract 228043, by the
Deutsche Forschungsgemeinschaft via the Research Unit FOR 538 and grant
GR~1132/15, and by the  German Excellence Initiative via the Nanosystems
Initiative Munich.

\section*{References}
\bibliographystyle{unsrt}

\end{document}